\documentclass[11pt]{article}
\usepackage{moriond,epsfig}

\usepackage{atlasphysics}
\usepackage{sparticles} 
\usepackage{xspace}

\usepackage{amsfonts}
\usepackage{amsmath}
\usepackage{amssymb}

\def\Journal#1#2#3#4{{#1} {\bf #2}, #3 (#4)}

\def\sst{\scriptscriptstyle}

\def\ra{\rightarrow}

\def\be{\begin{equation}}
\def\ee{\end{equation}}
\def\bea{\begin{eqnarray}}
\def\eea{\end{eqnarray}}

\newcommand{\meff}{$m_{eff}$}

\newcommand{\gl}{\ensuremath{\tilde{g}}\xspace}
\newcommand{\stau}{\ensuremath{\tilde{\tau}_1}\xspace}
\newcommand{\slepton}{\ensuremath{\tilde{\ell}_R}\xspace}

\newcommand{\bone}   {\tilde{b}^{}_{1}}
\newcommand{\tone}   {\tilde{t}^{}_{1}}
\newcommand{\neut}  {\tilde{\chi}^{0}_{1}}  

\newcommand{\Nfive}{\ensuremath{N_{5}}\xspace}
\newcommand{\Mmess}{\ensuremath{M_{\mathrm{mess}}}\xspace}
\newcommand{\Cgrav}{\ensuremath{C_{\mathrm{grav}}}\xspace}

\begin{document}
\vspace*{4cm}
\title{Searches for third generation SUSY in ATLAS}

\author{Antoine Marzin on behalf of the ATLAS Collaboration}

\address{University of Oklahoma, USA}

\maketitle\abstracts{
Recent results on searches for third generation supersymmetry carried out by the ATLAS 
collaboration with 2.05~fb$^{-1}$ of $\sqrt{s}$ = 7 TeV pp collisions recorded with the LHC in 2011 are reported. 
These analyses focus on search for gluino- and squark-mediated stau production, 
direct scalar bottom pair production and gluino-mediated sbottom and stop pair production.}

\section{Introduction}

Supersymmetry (SUSY)~\cite{Martin:2011} 
provides an extension of the Standard Model (SM) by introducing supersymmetric partners of the known 
bosons and fermions.  
In the framework of an $R$-parity conserving minimal supersymmetric extension of the SM (MSSM), 
SUSY particles are produced in pairs 
and the lightest supersymmetric particle (LSP) is stable, providing a possible candidate 
for dark matter. 
An important motivation for SUSY third generation searches is the fact that SUSY can naturally resolve the hierarchy problem,  
by preventing ``unnatural'' fine-tuning in the Higgs sector, provided that  superpartners of the top quark ($\tilde{t}$, stop) 
 have relatively low masses. This condition requires that the superpartner of the gluon ($\tilde{g}$, gluino) is not heavier than about 1.5 TeV 
due to its contribution to the radiative correction of the stop mass. 
Furthermore, in the MSSM the scalar partners 
of right-handed and left-handed fermions,
$\tilde{f}^{}_{R}$ and $\tilde{f}^{}_{L}$, can mix to form two mass eigenstates. This mixing 
is proportional to the corresponding SM fermion masses and is therefore more important for the third generation.
Large mixing can yield stau ($\tilde{\tau}_1$), sbottom ($\tilde{b}^{}_{1}$) and stop ($\tilde{t}^{}_{1}$) mass eigenstates 
which are significantly lighter than other sparticles. Consequently, they 
could be produced with large cross sections at the LHC.
Depending on the SUSY particle mass spectrum,
the cascade decays of gluino-mediated and pair-produced sbottoms
or stops result in complex final states consisting of missing transverse momentum (\met), several
jets, among which $b$-quark jets are expected, and possibly
leptons.

In this document, several ATLAS searches for third generation supersymmetry carried out  
using 2.05 fb$^{−1}$ of LHC pp data at $\sqrt{s}$ = 7 TeV
are reported. No significant excess above the SM expectation has been observed and exclusion  
limits at 95\% confidence level (C.L.) on SUSY parameters or masses of SUSY particles are derived.

\section{Search for gluino- and squark-mediated stau production} 

Two searches for events with large \met, at least two jets and at least 
one~\cite{1stau} or two~\cite{2stau} hadronically decaying tau leptons ($\tau$) have been carried out. The results have been
interpreted in the framework of minimal gauge mediated supersymmetry breaking (GMSB) 
which can be described by six parameters : the SUSY breaking mass
scale felt by the low-energy sector ($\Lambda$), the messenger mass
(\Mmess), the number of SU(5) messengers (\Nfive), the ratio of the
vacuum expectation values of the two Higgs doublets ($\tan\beta$), the
Higgs sector mixing parameter ($\mu$) and the scale factor for the
gravitino mass (\Cgrav). 
Assuming \Mmess = 250 TeV, \Nfive = 3, $\mu >$ 0 and \Cgrav = 1, 
squarks and/or gluino pairs are expected to be copiously produced at the LHC. 
These sparticles then decay directly or through cascades into the  next-to-lightest supersymmetric particle (NLSP),  
which subsequently decays into its SM partner and the LSP (light gravitino $\tilde{G}$). 
The experimental signature of the final state is thus driven by the nature of the LSP, 
which is the stau ($\tilde{\tau}_1$) for  a large part of the parameter space 
at large $\tan\beta$. 

The dominant backgrounds in the 1-$\tau$ analysis 
arise from top-pair plus single top production, vector boson production ($W$/$Z$ + jets) and multi-jets production, 
either with real $\tau$ leptons or mis-reconstructed $\tau$ from hadronic activity in the final state. 
These backgrounds are estimated in a semi-data-driven way, by normalizing 
the Monte Carlo (MC) event yield to the observed event yield in dedicated control regions, and then using the simulation 
to extrapolate into the signal region. 
In the 2-$\tau$ analysis, the main backgrounds  are from top-pair, single top and $W$ events
with one real $\tau$ lepton correctly reconstructed and one  mis-reconstructed $\tau$. 
Their contributions are also estimated using a control region enriched in top and $W$ events.  
The subdominant contribution due to the background from $Z \rightarrow \tau \tau$ events is extracted from simulations.

\begin{figure}[t]
\begin{center}
\includegraphics[width=0.48\textwidth]{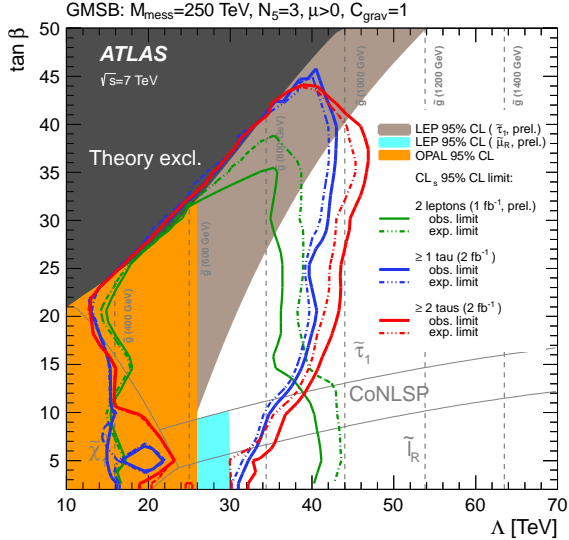}
\end{center}
\caption{Expected and observed exclusion limits at 95\%~C.L. on the
minimal GMSB model parameters $\Lambda$ and $\tan\beta$ assuming \Mmess = 250 TeV,
\Nfive = 3, $\mu >$ 0 and \Cgrav = 1.
The dark grey area indicates the region which is theoretically excluded due to unphysical sparticle mass values.
The different NLSP regions are indicated. In the CoNLSP region the \stau and the \slepton are the NLSP.}
  \label{fig:gmsb:limit}
\end{figure}

Figure~\ref{fig:gmsb:limit} shows the expected and observed exclusion limits at 95\% C.L. on $\Lambda$ and
on $\tan\beta$ as derived with the one and two $\tau$ leptons analyses. 
These results significantly improve 
the exclusion limits obtained with the previous ATLAS search in two opposite-sign leptons and the LEP results.

\section{Search for direct sbottom pair production} 

As search~\cite{mct} for direct sbottom pair production has been performed 
assuming  sbottom decay into a bottom quark plus a neutralino (LSP) with a branching ratio 
100\%. Selected events are required to have exactly two $b$-tagged jets with  
$p_T >$ 130, 50~GeV and \met\ $>$ 130~GeV.  Electrons (muons) with $p_T >$~20~GeV (10~GeV)
are vetoed, and events are rejected if a third jet with $p_T > $~50~GeV is found. The cuts on the leading jet and the \met\ 
are driven by the trigger thresholds. The kinematic variable used 
to further discriminate the signal from the background 
is the boosted-corrected contransverse mass $m_{CT}$~\cite{Tovey:2008ui,Polesello:2009rn}. 
The contransverse mass for two pair produced heavy particles 
with semi-invisible decay is defined as 
$([E_T (v_1) + E_T (v_2) ]^2 - [\mathbf{p_T} (v_1) - \mathbf{p_T} (v_2)])^{1/2}$, 
where $v_1$ and $v_2$ are the visible products of each decay chain. 
In the case of the considered signal, the $m_{CT}$ distribution has an end-point at
   $[m(\tilde{b}_1)^2 - m(\tilde{\chi}_1^0)^2]/m(\tilde{b}_1)$. 
The boosted-corrected contransverse mass is corrected for recoils in the 
transverse plane against initial state radiation to preserve the expected 
end-point in the distribution. 
Three signal regions are defined with $m_{CT} >$~100, 150 and 200~GeV  
to maximize the sensitivity for different mass splitting between the sbottom and 
the neutralino. 

The dominant SM background processes are top-pair plus single top production and asso\-ciated production 
of $W$/$Z$ bosons with heavy flavour jets. 
The \ttbar\ background is dominant in the signal region with $m_{CT} >$ 100~GeV due to 
the end-point at 135~GeV in the $m_{CT}$ distribution for \ttbar\ events. The two tighter 
signal regions are dominated by $Z \rightarrow \nu \nu$ + heavy flavour events, followed 
by  $W \rightarrow \tau \nu$ + heavy flavour events.
The sum of the top and $W$ plus heavy flavour contributions is estimated in a 
1-lepton control region, while the contribution from $Z$ plus heavy flavour production is estimated in a 
2-leptons control region. 
The background yield in each signal region is then obtained by multiplying the number of events observed in the corresponding control region by 
 the transfer factors defined as the ratio of the MC predicted yield in the signal region 
to that in the control regions. Subdominant backgrounds from diboson, \ttbar\ + $b\bar{b}$ and   \ttbar\ + $W$/$Z$ are 
estimated from MC simulations. 
The contribution from multi-jet events with possibly large \met\ is obtained by smearing jet energies in low \met\ ``seed'' events according 
to jet response functions extracted from MC simulation and tuned to data. This prediction is then normalised to the observed event 
yield in a multi-jets dominated control region.

\begin{figure}[t]
\begin{center}
\includegraphics[width=0.48\textwidth]{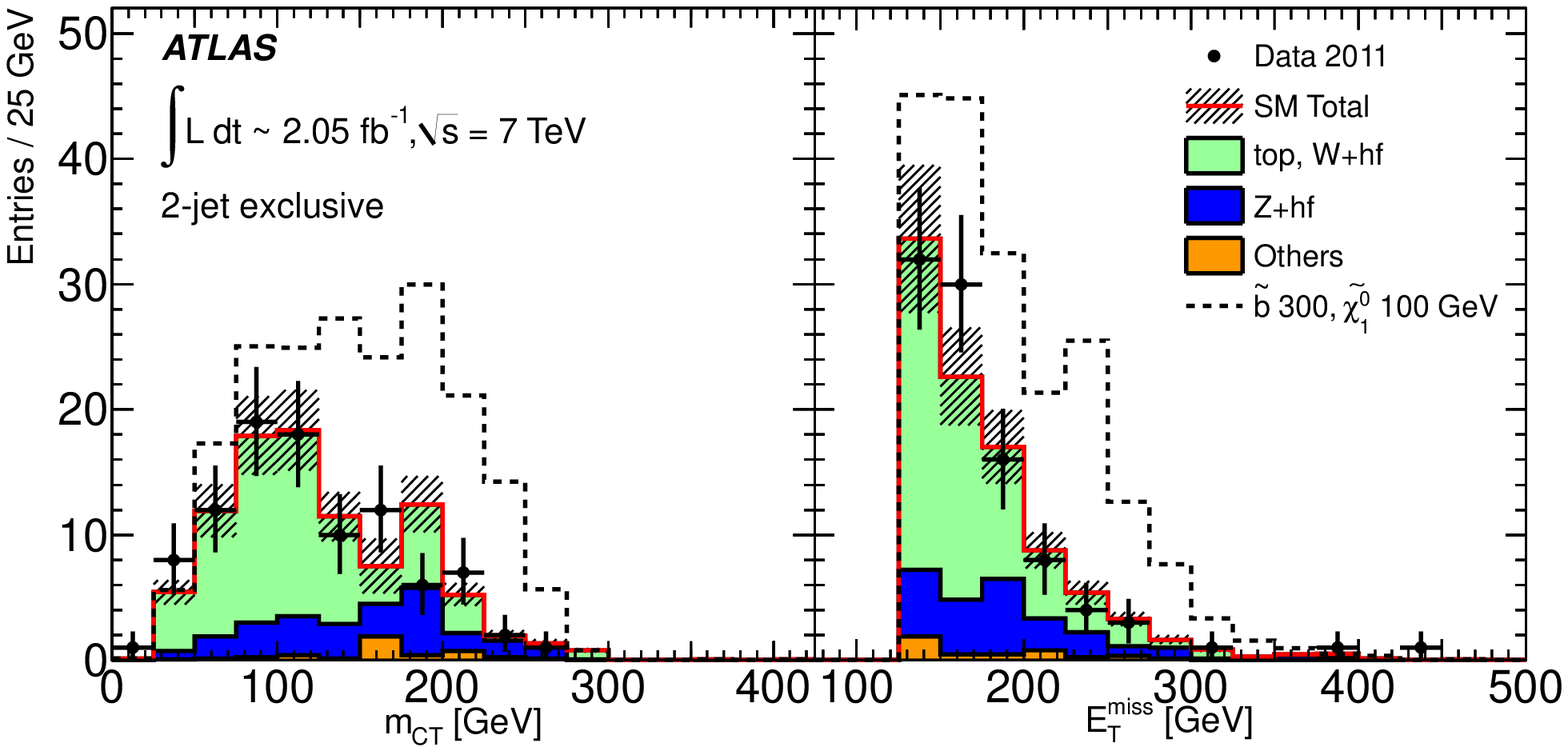} 
\includegraphics[width=0.48\textwidth]{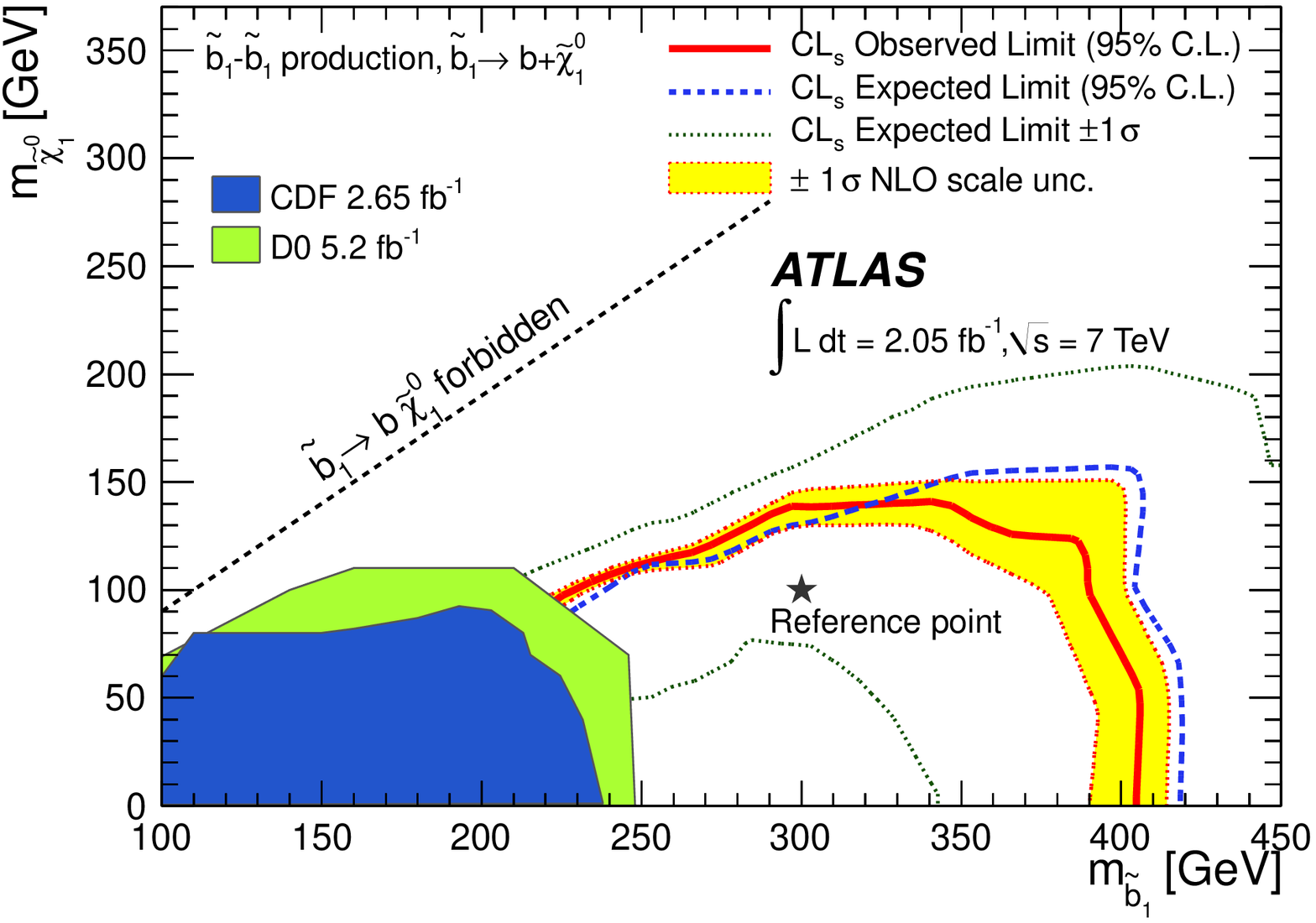}
\end{center}
\caption{{\itshape Left :} $m_{CT}$ and \met\ distributions before the final $m_{CT}$ cuts. {\itshape Right :} 
Expected and observed 95\%~C.L. exclusion limits  in the ($m_{\bone},m_{\neut}$) plane
resulting from the analysis searching for sbottom quark pair production, assuming sbottom to bottom plus neutralino decay.}
  \label{fig:sb:limit}
\end{figure}

Figure~\ref{fig:sb:limit} ({\itshape left}) shows the $m_{CT}$ and \met\ distributions before the final $m_{CT}$ cuts for data 
and SM background as predicted by the semi-data-driven method.  Figure~\ref{fig:sb:limit} ({\itshape right}) shows the 
expected and observed exclusion limits at 95\% C.L. in the  ($m_{\bone},m_{\neut}$) plane. For each signal point, the signal region leading to the best expected limit is chosen to extract the exclusion limits.
In the most conservative hypothesis, sbottom masses up to 390~GeV are excluded for neutralino masses below 60~GeV, which  significantly extends 
the previous results from the CDF and D0 experiments.

\section{Search for gluino-mediated sbottom pair production} 

ATLAS also searched~\cite{bjet} for gluino-mediated sbottom pair production 
assuming on-shell or off-shell sbottom decay into a bottom quark plus a neutralino (LSP) with a branching ratio of 
100\%, leading to a final state of four $b$-jets plus \met. In this analysis, events are selected by requiring at least three jets with 
$p_T >$ 130, 50, 50~GeV, \met\ $>$ 130~GeV and no lepton, the thresholds on the leading jet and \met\ 
being driven by the trigger requirement. Six signal regions are then characterized 
by the number of $b$-tagged jets ($\geqslant$ 1, 2) and the cut on the 
 effective mass \meff\ ($>$~500, 700, 900~GeV) defined as the scalar $p_T$
sum of all selected objects in the event. 

The strategy employed to estimate the SM backgrounds is similar to the 
semi-data-driven method used in the direct sbottom search. The dominant top background is estimated 
using two control regions which differ only in the number of $b$-jets required. 
These control regions are defined by applying the same selection cuts as for the signal regions, 
but requiring exactly 1 isolated lepton ($e$,$\mu$). The multi-jets background is estimated with 
the jet smearing method described above and the remaining contribution from 
$W$ and $Z$ production in association with heavy flavour jets is estimated using MC simulations. 

\begin{figure}[t]
\begin{center}
\includegraphics[width=0.48\textwidth]{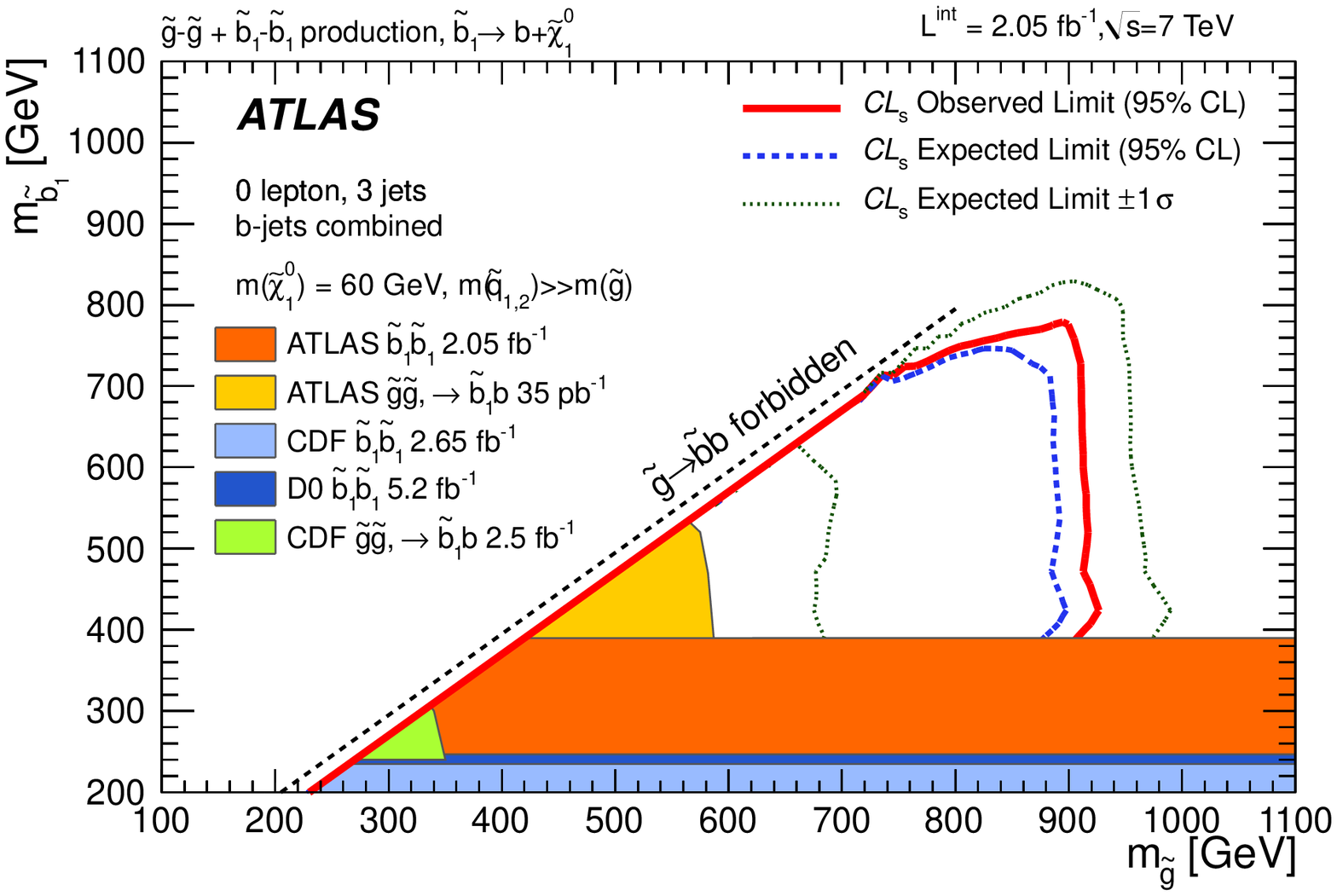} 
\includegraphics[width=0.48\textwidth]{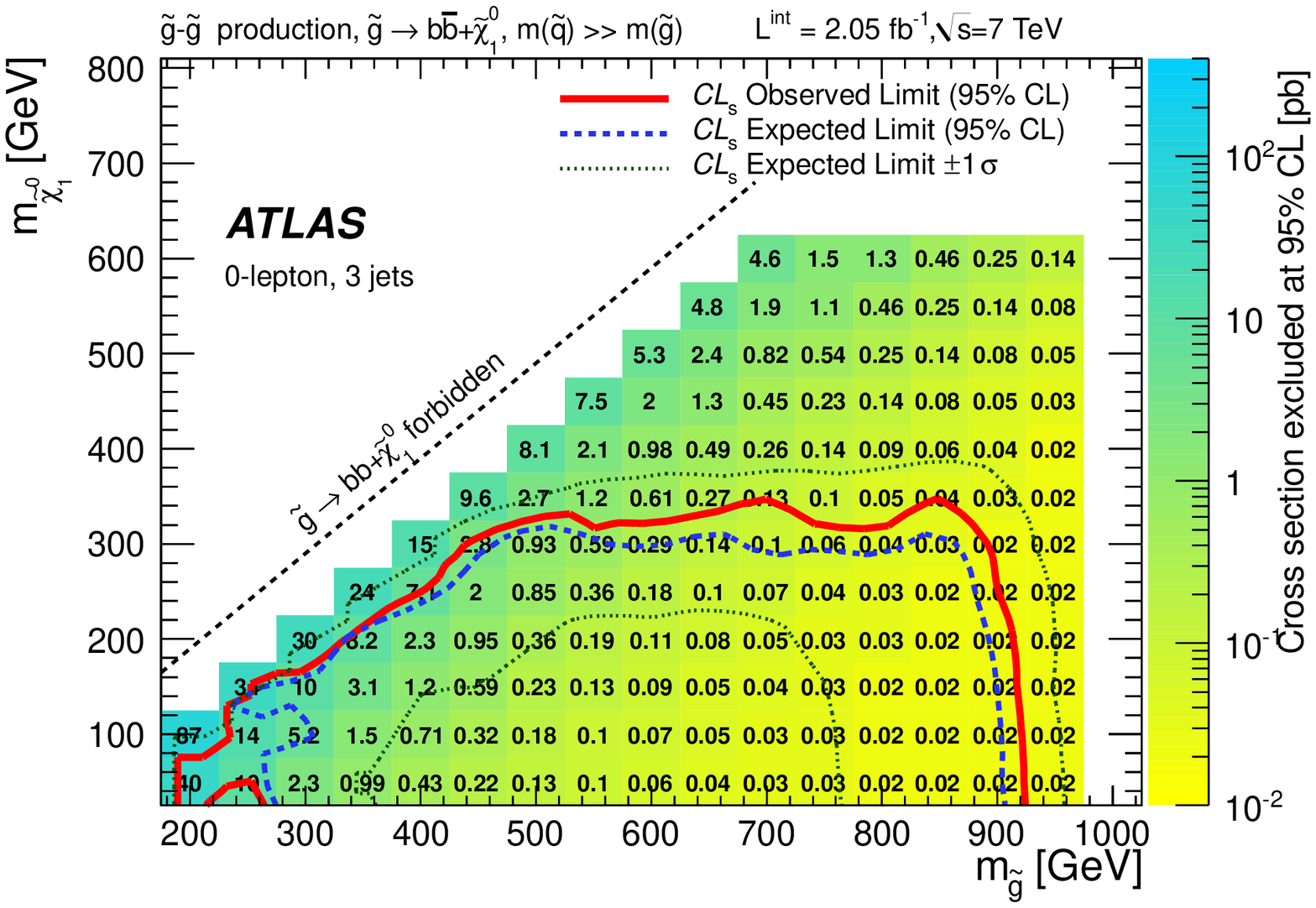} 
\end{center}
\caption{{\itshape Left :} Expected and observed 95\%~C.L. exclusion limits in the context of a
 MSSM model  in the ($m_{\tilde{g}},m_{\tilde{b}_1}$) plane. 
{\itshape Right :} Expected and observed 95\%~C.L. exclusion limits in the context of a simplified
model  in the ($m_{\tilde{g}},m_{\tilde{\chi}_1^0}$)  plane.}
  \label{fig:sbgl:limit}
\end{figure}

Results are first interpreted in the context of a  MSSM scenario where the $\tilde{b}_1$ is the 
lightest squark and all other squarks are heavier than the gluino  
(with $m^{}_{\gl}>m_{\tilde{b}_1} + m_b $) such that 
the branching ratio for $\tilde{g} \rightarrow  \tilde{b}_1 b$ is 100\%. 
In this case, the sbottom is produced in gluino decays of via direct pair production and 
is assumed to decay exclusively via 
$\tilde{b_1} \rightarrow b + \tilde{\chi}^{0}_{1}$, where the neutralino mass is 
set at 60~GeV.
The expected and observed 95\%~C.L. exclusion limits in the  ($m_{\tilde{g}},m_{\tilde{b}_1}$) plane are shown 
on Figure~\ref{fig:sbgl:limit} ({\itshape left}). Gluino masses
below 920~GeV are excluded for sbottom masses up to about 800~GeV.
 
Results are then interpreted in the context of a simplified model where 
the $\tilde{b}_1$ is the  lightest squark 
but with a mass above the TeV scale such that 
pair production of gluinos is the only process taken into account. 
A three-body final decay $\gl \rightarrow b\bar{b}\neut$ is assumed for the gluino with 
a branching ratio of 100\% via an off-shell sbottom decay $\tilde{b}_1^* \rightarrow b + \neut$.
Such a scenario, defined in a ($m^{}_{\gl},m^{}_{\neut}$) plane at fixed (large) sbottom mass,
can be considered complementary to the previous one, defined in the ($m^{}_{\gl},m^{}_{\bone}$)
  plane at fixed $\neut$ mass. 
The expected and observed 95\%~C.L. exclusion limits and the maximum 95\% C.L. upper cross section limits 
in the  ($m^{}_{\gl},m^{}_{\neut}$) plane are shown on Figure~\ref{fig:sbgl:limit} ({\itshape right}). 
Gluino masses below 900~GeV are excluded for neutralino masses below 300~GeV.

\section{Search for gluino-mediated stop pair production} 

Finally, two analyses targeting gluino-mediated stop pair production have been performed selecting 
events with one~\cite{bjet} or two~\cite{2lSS} isolated leptons ($e$,$\mu$). 
Assuming the  stop decays $\tilde{t}_1 \rightarrow t + \tilde{\chi}_1^0$ and $\tilde{t}_1 \rightarrow b + \tilde{\chi}_1^\pm$, 
the final state consists in many jets, including $b$-jets, leptons and large \met.  
In the one lepton analysis, events are selected if they contain exactly one isolated lepton and 
at least four jets with $p_T >$ 60, 50, 50, 50~GeV, amongst them a least one $b$-tagged jet. 
The transverse mass $m_T$ between the lepton and the \met\ must be larger than 100~GeV and 
the effective mass greater than 700~GeV. 
Two signal regions are then defined applying a cut on the \met\ at 80~GeV or 200~GeV.  
In the two leptons analysis, the selection requires at least two leptons 
with the same charge, at least four jets with $p_T >$ 50~GeV and  \met\ greater than 150~GeV. 
Two signal regions are then defined by applying or not a cut at 100~GeV on the 
 transverse mass $m_T$ between the highest $p_T$ lepton and the \met.

The SM background in the 1-lepton analysis is dominated by \ttbar\ events, followed 
by $W$ events produced in association with heavy flavour jets. 
In the 2-lepton analysis, the dominant SM background processes are $t\bar{t} W$, $t\bar{t} Z$ and 
$t\bar{t} WW$ (reffered as \ttbar +X), followed by multi-jets production with a non-prompt lepton  arising 
from $b$/$c$ decay, $\gamma$ conversion or jet misidentification. 
The number of multi-jets events  is estimated in both analyses 
using a matrix method based on the event count in two data samples which differ only 
by the lepton selection criteria. The contribution from other SM background sources 
in the 1-lepton analysis is normalized using the same semi-data-driven method as 
in the previous analyses. 
The control region is defined by reverting the $m_T$ cut and by relaxing the cuts 
on the $m_{eff}$ and the  \met\ to 500~GeV and 80~GeV, respectively, to increase the statistics and 
minimize the contamination from hypothetical signal events. 
The contribution of background events with charge misidentification 
in the 2-leptons analysis is estimated using a partially data-driven technique. 
This background is dominated by electron producing hard bremsstrahlung with subsequent photon conversion, 
the contribution from muon with incorrect charge assignment being negligible.  
The method consists in applying the probability of charge misidentification, measured in MC and tuned to 
data, to simulated \ttbar\ events with $e^\pm \ell^\mp$ in the final state. 
The other SM background processes, namely \ttbar + X and diboson, are extracted from simulations. 

\begin{figure}[t]
\begin{center}
\includegraphics[width=0.48\textwidth]{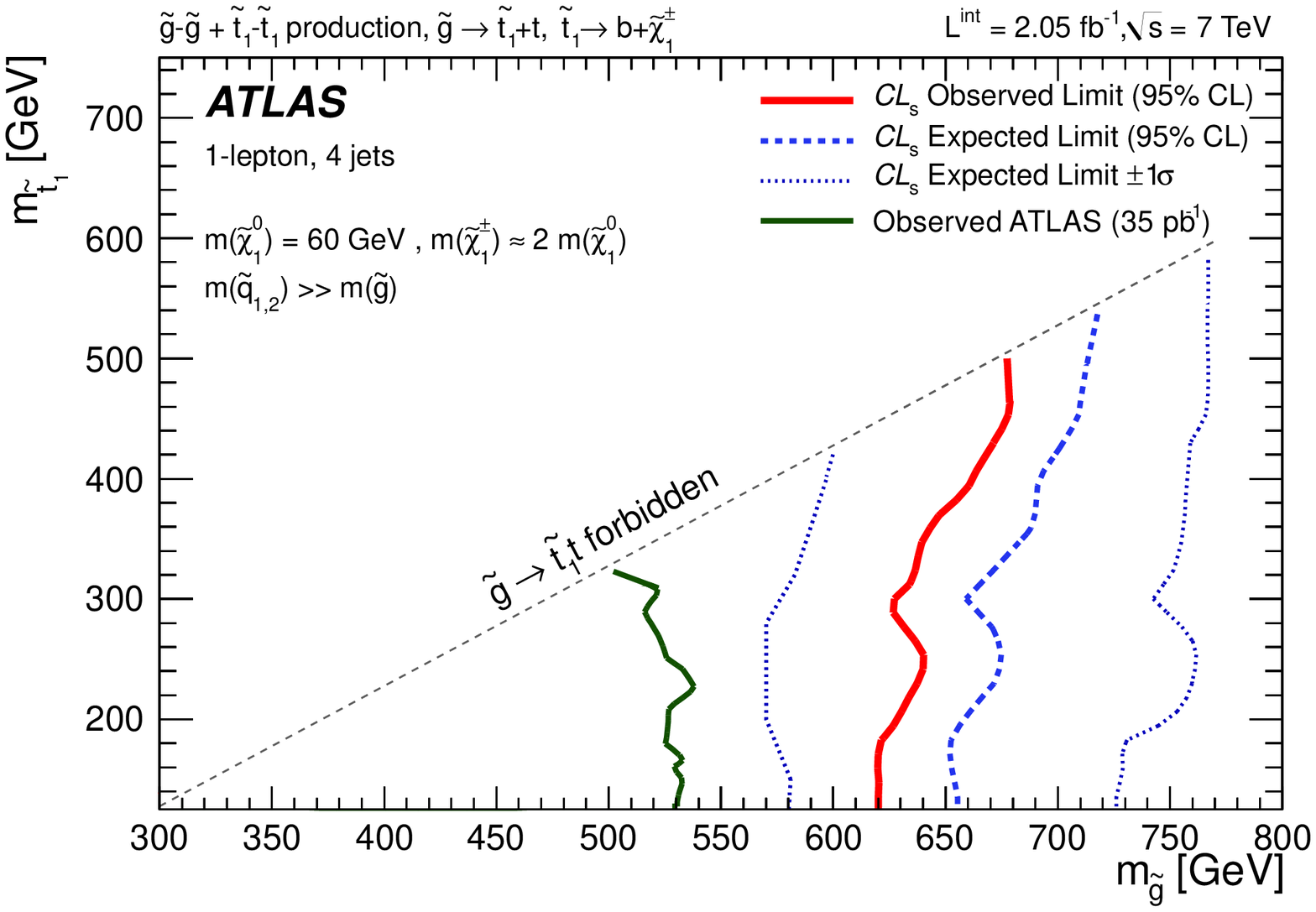} 
\includegraphics[width=0.48\textwidth]{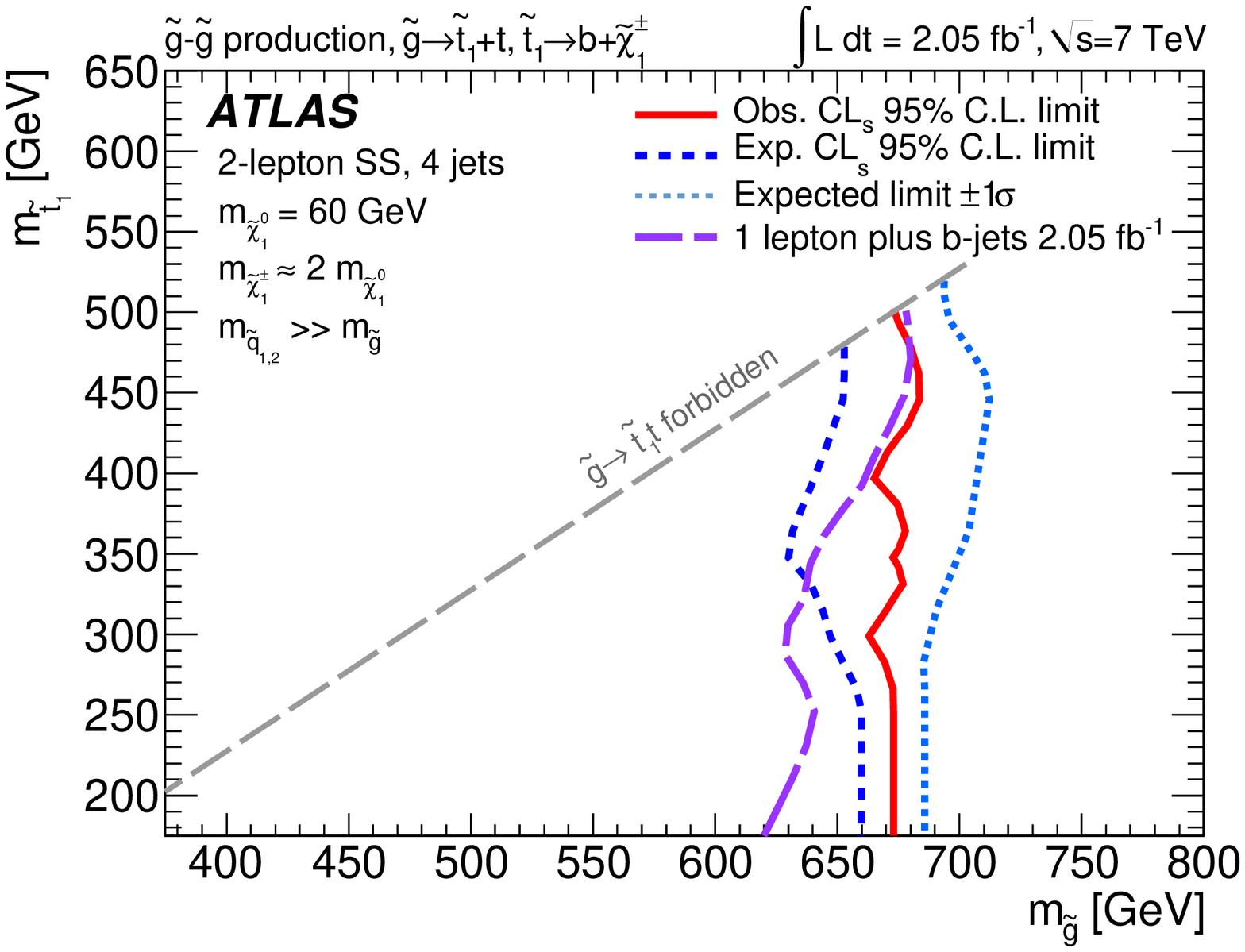} 
\end{center}
\caption{Expected and observed 95\%~C.L. exclusion limits in the context of a constrained
 MSSM model  in the ($m_{\tilde{g}},m_{\tilde{b}_1}$) plane as obtained with the 
1-lepton ({\itshape left}) and 2-leptons  ({\itshape Right}) analyses.} 
  \label{fig:stgl:limit}
\end{figure}

Results are first interpreted in the context of a  MSSM scenario where the $\tilde{t}_1$ is the 
lightest squark and all other squarks are heavier than the gluino  
(with $m^{}_{\gl} > m_{\tilde{t}_1} + m_t$ ) such that 
the branching ratio for $\tilde{g} \rightarrow  \tilde{t}_1 t$ is 100\%. 
In this case, the stop is produced in gluino decays of via direct pair production and 
is assumed to decay exclusively via 
$\tilde{t_1} \rightarrow b + \tilde{\chi}^{\pm}_{1}$, with the neutralino mass fixed  
 at 60~GeV. 
Figure~\ref{fig:stgl:limit}  shows the expected and observed 95\%~C.L. exclusion limits in the  ($m_{\tilde{g}},m_{\tilde{t}_1}$) plane 
 for both 1-lepton ({\itshape left}) and 2-lepton analyses  ({\itshape right}). Gluino masses
below 660~GeV are excluded for sbottom masses up to about 460~GeV.

\begin{figure}[t]
\begin{center}
\includegraphics[width=0.48\textwidth]{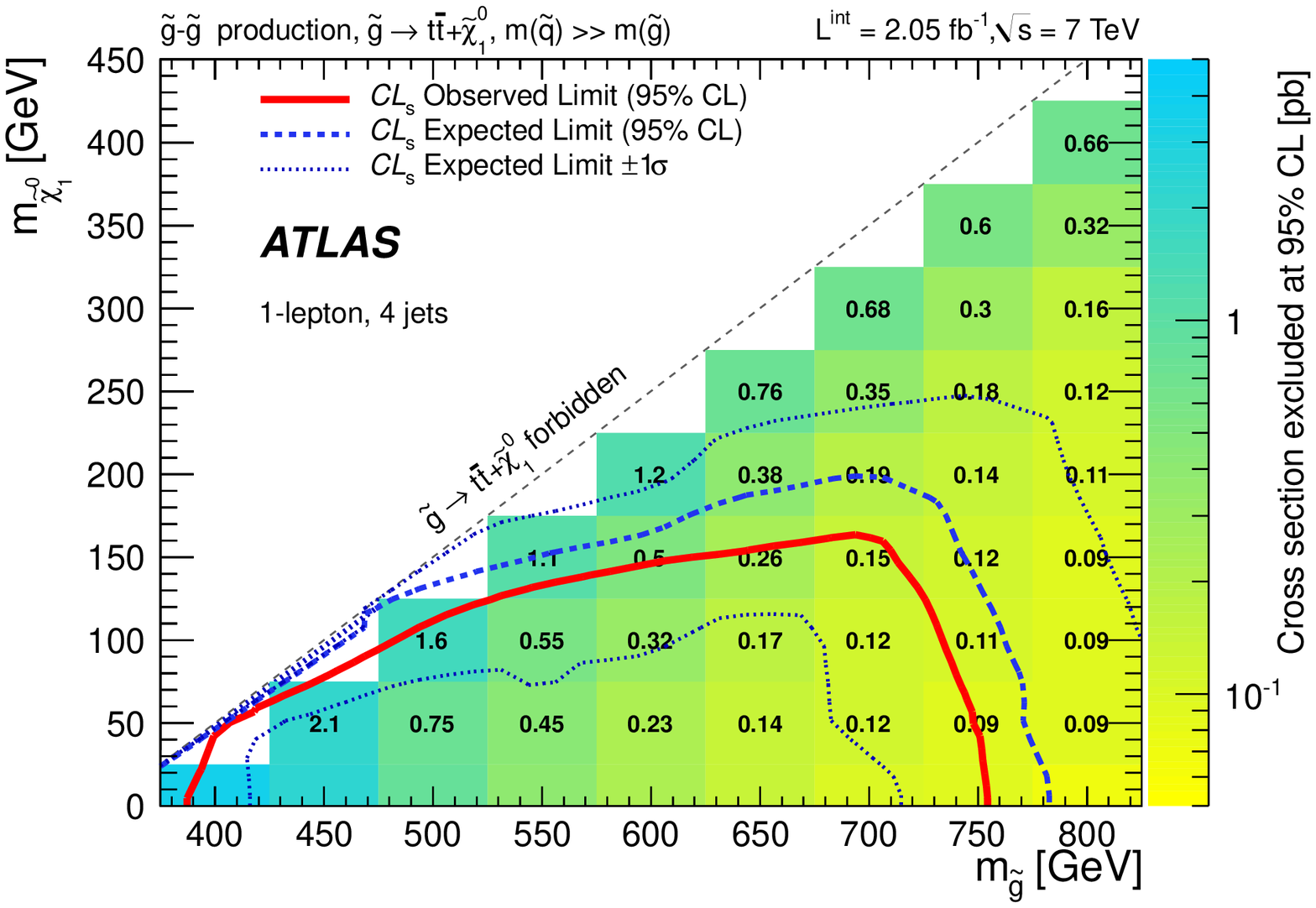} 
\includegraphics[width=0.48\textwidth]{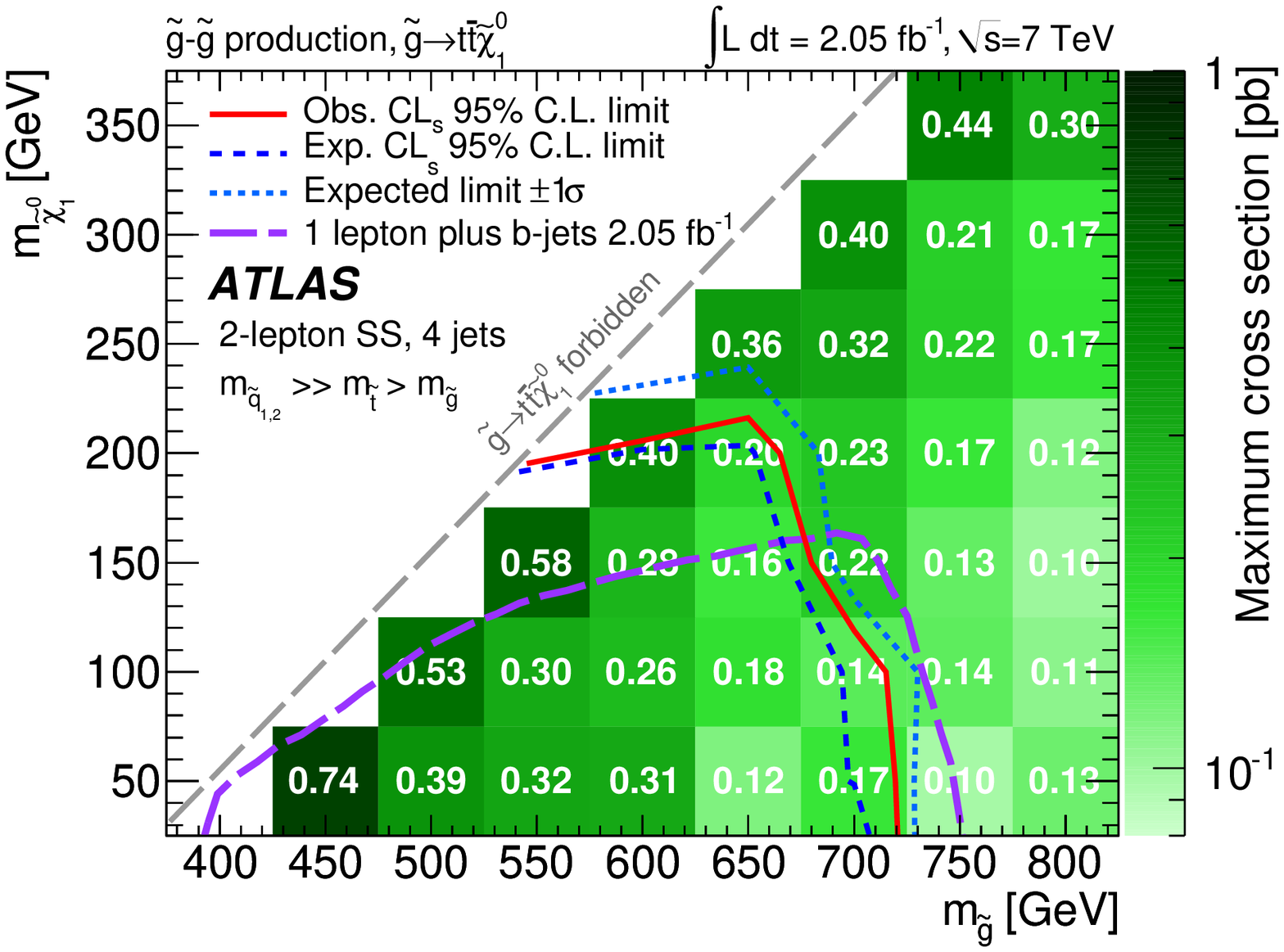} 
\end{center}
\caption{Expected and observed 95\%~CL exclusion limits in the context of a simplified
model  in the ($m_{\tilde{g}},m_{\tilde{\chi}_1^0}$) plane as obtained with the 
1-lepton ({\itshape left}) and 2-leptons ({\itshape right}) analyses.}
  \label{fig:Gtt:limit}
\end{figure}

Results are then interpreted in the context of a simplified model where 
the $\tilde{t}_1$ is the  lightest squark but with a mass above the TeV scale such that 
pair production of gluinos is the only process taken into account.  
A three-body final decay $\gl \rightarrow t\bar{t}\neut$ is assumed for the gluino with 
a branching ratio of 100\% via an off-shell stop decay $\tilde{t}_1^* \rightarrow t + \neut$.
Such a scenario, defined in a ($m^{}_{\gl},m^{}_{\neut}$) plane at fixed (large) stop mass,
can be considered complementary to the previous one, defined in the $m^{}_{\gl},m^{}_{\tone}$
mass plane at fixed $\neut$ mass. 
Resulting limits  in the ($m_{\tilde{g}},m_{\tilde{\chi}_1^0}$) plane 
are shown on Figure~\ref{fig:Gtt:limit} for the 1-lepton ({\itshape left}) and 2-leptons analyses  ({\itshape right}).
Gluino masses below 750~GeV are excluded for neutralino masses up to about 50~GeV. 
The 2-leptons analysis has the  best sensitivity at low $\tilde{t}_1 - \tilde{\chi}_1^0$ mass splitting 
due to softer kinematic cuts.

\section{Conclusion and prospects} 

ATLAS has carried out several searches 
 for superpartners of third generation fermions with an integrated luminosity of 2.05~fb$^{-1}$ . 
No excess in data with respect to the SM expectation has been observed so far.  
However, large regions of the parameter space for ``natural'' SUSY are still not excluded. 
In particular, no direct limits on the stop mass have been derived yet. 
Searches for direct stop pair production are currently in progress. 
These searches are challenging due to similarity with the \ttbar\ final state for  
low stop masses, and due to the low cross sections for higher stop mass values. 
In this respect, new results obtained with the full 2011 data set, corresponding to 4.7 fb$^{-1}$, 
and the 2012 data at $\sqrt{s} = 8$~TeV will be very important.

\end{document}